# Exposing Application Components as Web Services

Scott M. Walker, Alan Dearle, Graham N.C. Kirby & Stuart J. Norcross

*School of Computer Science, University of St Andrews,
St Andrews, Fife KY16 9SX, Scotland*

**Abstract**

This paper explores technology permitting arbitrary application components to be exposed for remote access from other software. Using this, the application and its constituent components can be written without concern for its distribution. Software running in different address spaces, on different machines, can perform operations on the remotely accessible components. This is of utility in the creation of distributed applications and in permitting tools such as debuggers, component browsers, observers or remote probes access to application components. Current middleware systems do not allow arbitrary exposure of application components: instead, the programmer is forced to decide statically which classes of component will support remote accessibility. In the work described here, arbitrary components of any class can be dynamically exposed via Web Services. Traditional Web Services are extended with a remote reference scheme. This extension permits application components to be invoked using either the traditional pass-by-value semantics supported by Web Services or pass-by-reference semantics. The latter permits the preservation of local call semantics across address space boundaries.

## 1. Introduction

This paper introduces RRT, a system that permits arbitrary components to be exposed for remote access from arbitrary software. It allows dynamic deployment, as Web Services, of components in a running application. Remote method calls can be performed on the exposed components from different address spaces, possibly on different machines. RRT has five notable features that differentiate it from typical Web Service technologies.

1. Specific, existing component instances rather than component classes are deployed as Web Services.
2. The programmer does not need to decide statically which component classes support remote access. Any component from any application, including previously compiled applications, can be deployed as a Web Service without the need to access or alter the application's source code.
3. A remote reference scheme, synergistic with standard Web Services infrastructure, provides pass-by-reference semantics, in addition to the pass-by-value semantics supported by Web Services.
4. Parameter passing mechanisms are flexible and may be dynamically controlled through policies. A deployed component can be called using either pass-by-reference or pass-by-value semantics on a per-call basis.
5. The system automatically deploys referenced components on demand.

These features provide several potential benefits. Distribution can be introduced into a non-distributed application with little effort. Unlike conventional middleware systems, the programmer can write an application without concern for its distribution and introduce distribution at a later time. Since any component can be made remotely accessible, changes to distribution boundaries do not require re-engineering of the application, making it easier to change the application's distribution topology.

The process of creating an application is separated from the process of distributing it. This separation of concerns simplifies the software engineering process to the programmer's advantage both when creating a distributed application and when introducing distribution to a previously non-distributed application.

A component within a non-distributed application may be made remotely accessible without the need to modify the application's source code. This simplifies the creation of tools such as debuggers or application probes that need to access component state from another address space. For example, the programmer can introduce a remote observer to a non-distributed observable component, and component browsers can attach remotely to any application, permitting object closures to be remotely browsed. Using traditional middleware systems, it is difficult to attach such tools to existing components without access to source code and extensive engineering effort.

Current approaches to creating distributed applications using middleware suffer from the following problems:

- They force decisions to be made early in the design process about which classes may participate in inter-machine communication.
- They are brittle with respect to changes in the way in which the application is distributed.
- It is difficult to understand and maintain the distributed application since particular parameter passing mechanisms may force an unnatural encoding of application level semantics.

In contrast, using the technology described in this paper, the application and the components do not need to have been written to explicitly support distribution. This provides the advantages of flexibility in application distribution and simplification of application implementation. Without this technology, the creation of tools such as probes, debuggers or observers that attach to precompiled components is difficult.

## 2. Related Work

Web Services provide an RPC mechanism. A Web Service is a remote interface to a component class that has been deployed in a Web Service container. The Web Service container acts as a web server accepting incoming method calls in the form of HTTP requests. The URL specified in the request indicates which Web Service is being invoked. The body of the request contains the name of the method to invoke and the arguments to be passed, encoded using SOAP [1]. The Web Service Description Language (WSDL) [2] is used to describe the methods available in a Web Service.

Web Service technologies such as Apache Axis [3] and Microsoft .NET Web Services [4] deploy a class of component, not a specific instance of a component. The class is automatically instantiated to handle incoming requests on a per-call basis or on first access. Web Services systems do not permit the deployment of a specific component. Consequently, using standard Web Services, the only way in which specific components can be accessed is to manually provide a multiplexing Web Service which maps from keys to specific components. This makes it difficult to expose application components using standard Web Service technology.

A Web Service can only use service components instantiated in its container's address space. The service component cannot access a component in a different address space. There is no general mechanism to run an existing application in a Web Service container. As a result, it is difficult to link a Web Service with an existing application.

Web Service technologies do not provide any form of remote object reference scheme. Web Services use only pass-by-value semantics. In contrast, Distributed Object Models (DOMs) provide both RPC mechanisms and remote object reference schemes but do not allow arbitrary exposure of application components. A reference to a remotely accessible component can be passed across address space boundaries. Method calls performed on the remotely referenced component are transparently propagated across the network to the referenced component.

The creation of a remotely accessible component using DOMs such as CORBA [5], Java RMI [6] and Microsoft .NET remoting [4] requires the programmer to follow similar steps:

- The programmer is forced to decide statically the interfaces between distribution boundaries.
- The programmer is forced to decide statically which classes of component will implement these interfaces and thus be remotely accessible.

- These remotely accessible classes must extend a special base class that provides the functionality necessary for remote accessibility[1]. This has two effects: to force the static identification of accessible classes, as above, and, in languages without multiple inheritance, to prevent the creation of accessible subclasses of existing non-accessible classes.
- Once a remotely accessible class is instantiated, the instance is associated with a naming service that allows remote callers to obtain a remote reference to it.

Some research DOMs, such as JavaParty [7] and ProActive [8], have similar motivation to the described work—to hide from the programmer the complexity of using a more conventional DOM. JavaParty automatically generates RMI-based source code, based on the introduction of a new Java keyword (`remote`). The programmer must still statically decide which classes of component support remote accessibility. ProActive is a Java library that provides tools for the creation of distributed applications. The ProActive work is closest to the work described in this paper. It differs in that only certain *active objects* may be remotely accessed and sharing of passive objects is forbidden. In ProActive, each active object carries out the work performed by RRT in our system. By contrast, in our work there is only one RRT instance per address space, which provides a view onto arbitrary application components. Furthermore, ProActive is based on RMI and JMS whereas RRT is based on Web Services.

In existing DOMs, the transmission policy for arguments and return values, that is, whether they are passed across the network by value or by reference, is statically fixed. Remotely accessible components are passed by reference, while all other components are passed by value. In contrast, in RRT, remotely accessible components may be passed either by reference or by value, and this choice may be made dynamically; this is described further in Section 5.

## 3. Deploying Components as Web Services

RRT was designed as a DOM that would support the arbitrary redistribution of an application. It has been developed as part of the RAFDA project [9]. RRT has been implemented using Java and all code examples in this paper are shown in this language.

RRT permits the dynamic exposure of arbitrary components as Web Services. Web Services were chosen as the implementation platform for several reasons. Firstly, they provide a standard call and return RPC mechanism that can be easily type checked. Secondly, Web Services provide cross-language, cross-platform interoperability.

To deploy a component, the programmer registers it with RRT, specifying a Java interface that describes which of the component's methods are to be remotely accessible. Only these methods then appear in the corresponding Web Service. Note, however, that the component's class need not implement the specified Java interface—if this was necessary then the programmer would be forced to decide statically which classes could be remotely accessible, contrary to our design aims. In RRT the deployed component need only be of a class that *could* implement the interface, i.e. it contains equivalent methods to those specified in the interface. The importance of this is that the interface can be written when the deployment decision is made, potentially after the component already exists in a running application. Compatibility of the deployed component with the specified interface is checked at deployment.

Figure 1 shows a *Student* component being deployed using the *INamedEntity* interface—which it does not implement.

---

[1] This is not strictly true in CORBA implementations, which can use the *tie* approach to bind a servant to a POA. Our approach is similar.

```java
public class Student {
      private String name = null;
      private int matricNumber = 0;
      public Student(String name, int matricNumber) {
             this.name = name; this.matricNumber = matricNumber;
      }
      public String getName() {return this.name;}
      public int getMatriculationNumber() {return this.matricNumber;}
}

public interface INamedEntity {
      String getName();
}

Student aStudent = new Student("Bobby Jones", 1234);
RRT.deploy(INamedEntity.class, aStudent, "bob");
```

**Figure 1: Deploying a component**

Executing this code causes a new Web Service to be deployed that uses the *Student* component *aStudent* as its service component and exposes an *INamedEntity* interface[2]. The Web Service is located at a URL based on the specified name. In this case specifying "bob" results in the *Student* service component being exposed at the address:

    `http://hostname:port/bob`

where the hostname and port refer to the network interface and port on which RRT is running. As in standard Web Services, the corresponding WSDL can be obtained at the URL:

    `http://hostname:port/bob?wsdl`

When a request is sent to this URL, RRT returns the WSDL corresponding to the *INamedEntity* Java interface, which is automatically generated on demand. From the WSDL, conventional generative tools can be used to create a client application that accesses the deployed component.

In addition to generating WSDL, RRT generates a skeleton which is used to interface between RRT and the service component. This serves a similar purpose to skeletons in CORBA implementations, providing RRT with a well-known *invocation()* interface by which the service component may be invoked. The skeleton's *invocation()* method takes the name of the method to call and the arguments, which have been deserialised by RRT, and calls the specified method on the deployed component. The use of skeletons obviates the need for an expensive reflective operation during a method call from a remote address space, making it efficient at the cost of a one-time generative step on the first deployment of a particular interface.

The *RRT.deploy()* call would fail if the deployed component were not compatible with the specified interface.

### 4.   Extending the Web Services Paradigm to Support Remote References

When converting an existing non-distributed application into a distributed application, pass-by-reference semantics are desirable. Consider the following example:

---

[2] It is also possible to deploy a component multiple times with different interfaces.

```java
public interface IPerson {
        IPerson getSpouse();
        void setSpouse(IPerson spouse);
        int getAge();
        void incrementAge();
}

IPerson mary = findLocalOrRemotePerson("Mary Smith");
IPerson john = new Person("John Brown", 35);

mary.setSpouse(john);
john.incrementAge();
print(mary.getSpouse().getAge());
```

**Figure 2: Example of effect of parameter passing semantics**

When this program is executed, the output will be different depending on whether the component denoted by *mary* is local or remote. If it is local, the output will be 36, otherwise it will be 35. In the latter case, the parameter to *setSpouse()* will be passed by value, creating two copies of the *john* component in different address spaces, and the remote copy will not be updated. This example clearly illustrates that without pass-by-reference the application semantics depend on the distribution boundaries.

To address this need, RRT supports both pass-by-value and pass-by-reference semantics. To implement the latter, a mechanism is provided to allow remote references to be passed across Web Service interfaces. RRT represents a remote reference using a data structure that contains information describing the referenced component. Termed a RAFDA-IOR (Interoperable Object Reference), this contains the following information:

- A local object number that identifies the component uniquely within its RRT instance.
- The interface exposed by the referenced component.
- The host on which the RRT instance deploying the referenced component is running.
- The port at which the RRT instance is accessed.

When a RAFDA-IOR is exported by an RRT instance, a local object number is lazily assigned and an associative mapping from the local object number to the service component is created. This permits the service component to be discovered on reception of a RAFDA-IOR during an incoming remote call.

Since client side Web Service tools such as WSDL2Java provided in Apache Axis or the *wsdl.exe* tool provided in the Microsoft .NET Framework do not understand RAFDA-IORs, some additional client side tools are needed to exploit this extra functionality. These are provided by a client side instance of RRT. An important function provided by RRT when used in this role is to obtain a reference in the form of a RAFDA-IOR to a component that has been deployed in another RRT instance. This is achieved using the *getComponentByName()* method as shown in the following code fragment.

```java
INamedEntity remoteEntity = (INamedEntity)
        RRT.getComponentByName("bob", "koala.rafda.org", 5001);
```

**Figure 3: Obtaining a remote reference**

Figure 3 shows how a reference to the remote component exposed as an *INamedEntity* interface in Figure 1 may be obtained, assuming that it was deployed on host `koala.rafda.org` at port 5001. *The getComponentByName()* method is a wrapper method which receives a RAFDA-IOR from the remote site and generates a proxy component in the local address space, which it returns typed as *Object*. The proxy implements the same Java interface as the one with which the named component was deployed in the remote address space, in this case the *INamedEntity* interface. This process is followed whenever a RAFDA-IOR is received from a remote host. Conversely, whenever a proxy is passed another proxy, it is translated into either a RAFDA-IOR or a standard SOAP encoding, dependent on transmission policy (described in the next Section).

Consider again the example shown in Figure 2. If the identifier *mary* refers to a remote component, the local parameter *john* must be passed by reference during the *mary.setSpouse()* call. As discussed, this is achieved by encoding the reference to *john* as a RAFDA-IOR. This RAFDA-IOR is obtained by automatically deploying *john*, which follows the deployment process described above. Automatic deployment makes use of an anonymous *deploy()* method which is like the one shown in Figure 1 but does not take the last parameter naming the object. This Just-In-Time deployment relieves the programmer of the responsibility of ensuring consistent remote accessibility.

**5.     Caveats and Restrictions**

RRT has several limitations with respect to remote references using pass-by-reference semantics. Since distribution boundaries are flexible, a client holding a reference cannot know whether it is directly referencing an application component or a proxy. When using Java, in order to ensure that the reference is typed in such a way that either component or proxy can be referenced as required, all references must be typed as interfaces. In the example of Figure 2, the identifier *mary* is typed as an *IPerson* in order that either a local *Person* component or a *Person* proxy can be referenced transparently. To ensure this, the interface specified when deploying a component must itself refer only to interface types, rather than concrete classes. This rule applies recursively to all referenced interfaces. This property is checked during the *RRT.deploy()* call.

If components of a single class are distributed among several RRT instances then the semantics of static fields and methods will not match the non-distributed application. Static state will be shared only among the components in the same RRT instance.

Clearly these restrictions limit the degree to which an arbitrary application can be redistributed using RRT, since many applications will have been originally written using references typed as concrete classes, or using static fields. To address this problem, we have developed a set of transformational tools [10] that circumvent both restrictions. These operate at the Java byte code level; they take an arbitrary non-distributed application and transform it into a distributable version. They extract interfaces for all components and separate the static and instance parts of each component. The tools also handle code that is rendered non-transformable due to native methods and the special treatment of arrays as Objects in the Java Virtual Machine.

The transformed application generated using these tools uses factories for all component creation. Factories are capable of instantiating components in a remote RRT instance. A placement policy framework has been developed that allows the definition of placement policy of arbitrary complexity.

For compatibility with standard Web Services, the default parameter passing semantics in RRT is pass-by-value. However, as discussed above, in order to facilitate the construction of distributed applications that preserve single address space semantics, pass-by-reference is also supported. Unlike most DOMs, which force decisions on parameter passing semantics to be made early in the design cycle, RRT provides a flexible transmission policy framework, permitting the programmer to specify whether arguments should be passed by reference or by value. Furthermore, this policy can be applied to classes, methods and arguments on a system-wide or per-call basis, to provide maximum flexibility.

Figure 4 shows examples of transmission policies being applied to the examples shown in Figures 1 and 2. The call to *setClassPolicy()* illustrates how the policy manager may be used to specify pass-by-reference on a class-wide basis. Once such a call has been made, all instances of class *Student* will be passed by reference unless otherwise specified by another more specific policy. An example of such a more specific policy is shown in the call to *setMethodPolicy()*, which specifies that all parameters to the *IPerson.setSpouse()* method should be passed by value.

```
TransmissionPolicyManager.setClassPolicy(  Student.class,
                                          Policy.BY_REFERENCE);
TransmissionPolicyManager.setMethodPolicy( IPerson.class,
                                          "setSpouse",
                                          Policy.BY_VALUE);
```

Figure 4: Transmission Policy Examples

Other control methods include *setParamPolicy()*, which controls policy for individual parameters of a method, and *setReturnValuePolicy()*. The RRT transmission policy control mechanisms provide the expert programmer with fine-grained control over the manner in which parameters are passed, in line with the RRT design goals.

**6.    Conclusions**

RRT permits arbitrary components from arbitrary applications to be exposed for remote access as Web Services. It is not necessary that these components were originally written to support remote accessibility, in contrast to conventional middleware systems. Components can be accessed from other address spaces. This aids in the creation of distributed applications and the conversion of non-distributed applications into distributed ones. Tools that access application state from another address space can be created without modification, or even access, to the application's code.

RRT employs a remote reference scheme that operates in conjunction with the deployed Web Services to provide pass-by-reference semantics. This functionality permits the preservation of local call semantics across address space boundaries. Remote method call can be performed using either the traditional pass-by-value semantics supported by Web Services or pass-by-reference semantics.